\documentstyle[11pt,epsf]{article}
\begin{document}
\setlength{\baselineskip}{0.6cm}
\title{Investigations of Process Damping Forces in Metal Cutting}
\author{Emily Stone$^*$ \& Suhail Ahmed\\
{\it Department of Mathematics and Statistics}\\
{\it Utah State University}\\
{\it Logan, UT 84322-3900}
\\
and\\
Abe Askari \& Hong Tat\\
{\it Mathematics and Computing Technology}\\
{\it The Boeing Company}\\
{\it P.O. Box 3707 MS 7L-25}\\
{\it Seattle, WA 98124-2207}}

\maketitle
\vspace*{4.5in}
{\small
\noindent
{\bf * Communicating author. Current address: Dept. of Mathematical Sciences,
The University of Montana, Missoula, MT 59812}

\noindent
Running title:   Process Damping Forces in Metal Cutting}

\newpage
\begin{abstract}
Using finite element software developed for metal cutting
by Third Wave Systems we investigate the forces involved
in chatter, a self-sustained oscillation of the cutting tool.
The phenomena is decomposed into a vibrating tool cutting 
a flat surface work piece, and motionless tool cutting a work piece with
a wavy surface.  While cutting the wavy surface, the
shearplane was seen to oscillate in advance of the oscillation
of the depth of cut, as were the cutting, thrust, and shear plane forces.
The vibrating tool
was used to investigate process damping through the interaction
of the relief face of the tool and the workpiece.  Crushing 
forces are isolated and compared to the contact length between
the tool and workpiece.  We found that the wavelength dependence
of the forces depended on the relative size of the wavelength
to the length of the relief face of the tool.   The results 
indicate that the damping force from crushing will be proportional to the
cutting speed for short tools, and inversely proportional for
long tools.

{\bf keywords}: metal cutting, chatter, process damping, shear plane
dynamics.

{\bf AMS classification numbers:} 74S05,74H45,74H15.

\end{abstract}
\newpage

\section{Introduction}
In this paper we investigate the forces produced in dynamic
metal cutting, where either the chip load varies in time, or
the tool itself oscillates.  Chatter is a consequence of an
instability in cutting that leads to self-sustained vibrations
of the tool, hence it is composed of both an oscillating tool
and an undulating workpiece surface.  We separate chatter
into these two pieces by using finite element software for metal
cutting developed by Third Wave Systems, Inc. called AdvantEdge,
to simulate the processes.
AdvantEdge is a validated software package that integrates advanced
dynamic, thermo-mechanically coupled finite element numerics and material
modeling appropriate for machining processes.  The simulation software
provides accurate estimates of thermo-mechanical properties of the
machining process such as cutting forces, chip morphology, machined
surface residual stresses, and temperature behavior of the tool and workpiece.

Recently Third Wave Systems has developed two new modules, one with custom work
piece geometry and another with a vibrating tool, that increase the range of
cutting conditions that can be simulated.  The custom work piece
geometry module allows for the design of arbitrary cutting surfaces
in addition to the traditional flat surface. We study the effect
of a varying chip load through a cut with this module, observing 
interesting shear plane dynamics and varying thrust and cutting forces.  
With the vibrating tool module, the cutting tool can be horizontally
or vertically vibrated at a user specified frequency and amplitude.
This allows us to investigate the effect of contact between the material
and the relief face of the cutter, which generates a kind of process damping. 

In section 2 we examine the forces involved in  cutting a workpiece with a
sinusoidally varying surface and, as a consequense, a sinusoidally
varying depth of cut.  In section 3 we use the vibrating tool module
to investigate process damping on the relief face of the tool during
chatter.  In section 4 we summarize our findings and suggest avenues
for further research.

\section{Wavy Surface Runs}

The work piece
geometry we use has a sinusoidal top surface which was generated
by replacing the horizontal flat surface of the standard geometry
with a sine function.  The wavelength of the sinusoid was 
 based on scaled parameters from experimental chatter measurements \cite{STONE}.
  To ensure the feed was always greater than zero, the feed was
set to be larger than the peak to valley amplitude of the sinusoid (3 mil).
All these AdvantEdge runs were performed on AL 7050 with a carbide tool with
a 10 degree rake angle, and cutting edge radius of 0.7874 mil and cutting 
speed of 2680 SFM.
Altogether three sets of runs were performed with three different feeds: 6,9, and 12 mil.

From the animations of the Third Wave runs we saw that the
shear plane angle during the cut oscillates as the tool moves through 
sinusoidally varying depth of cut (see figure \ref{frames}). 
The shear plane, denoted by the band of higher heat rates extending
from the tool tip, is seen to lift up smoothly and shorten as the tool anticipates  
the thickest part of the chip.  After reaching its maximum slope, the band
snaps down and elongates to connect the tool tip with the point ahead on
the surface that has the smallest chip thickness.
The shear plane continues to extend between the tool tip and this minimum
until that point, now a part of the chip, is crushed into the point on
the surface with maximum chip thickness.  The cycle repeats as
the tool continues to encounter the oscillating chip thickness.
The experiments we performed were designed to explore the effect of variation
in depth of cut, shear plane angle, and shear plane length on the
instantaneous cutting forces.

\subsection{Wavy Surface Force Predictions}
\begin{figure}
\centerline{\epsffile{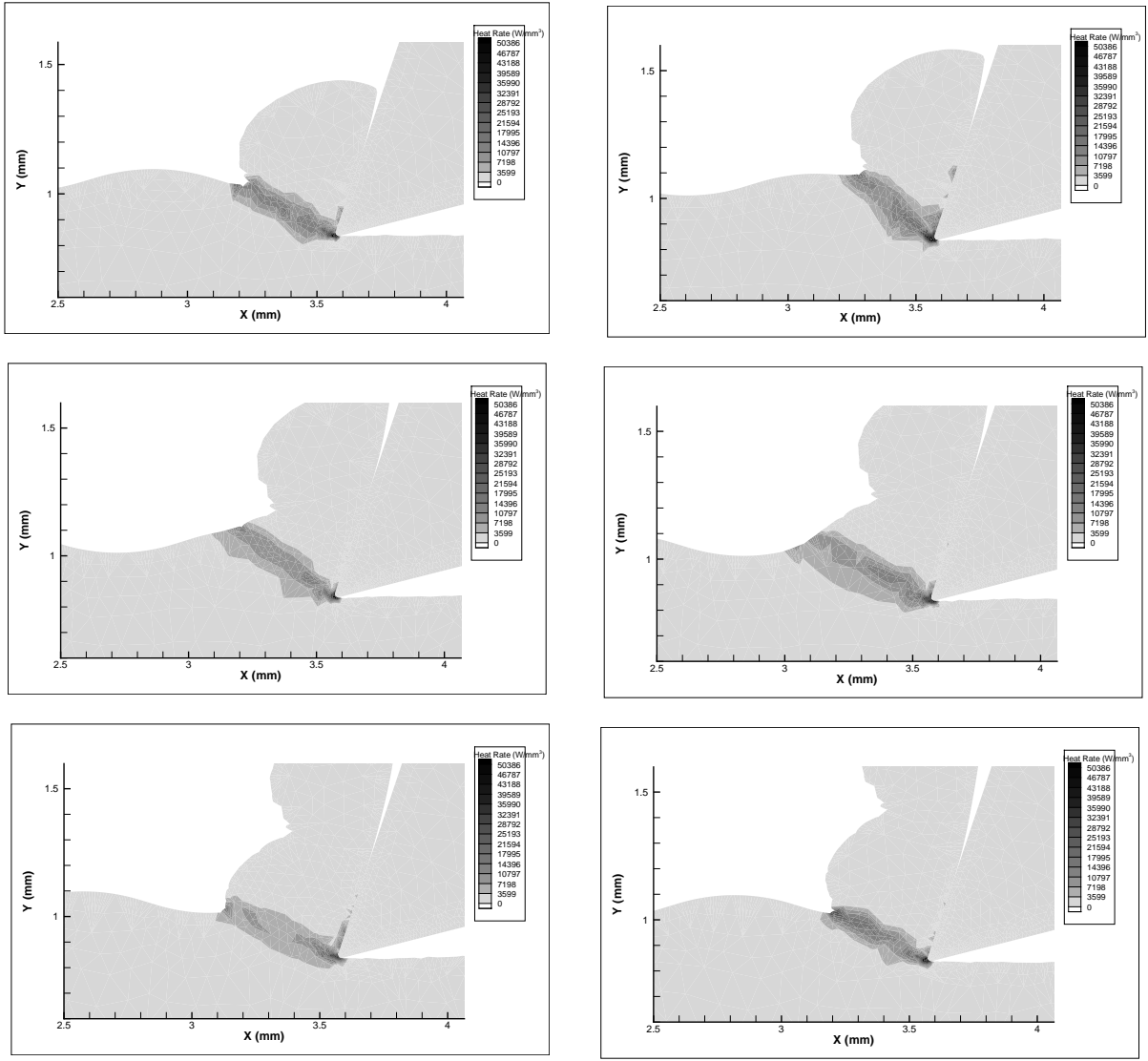}}
\caption{Sequence of snapshots from a Third Wave run, time increases to
the right and down.}
\label{frames}
\end{figure}

\begin{figure}
\centerline{\epsffile{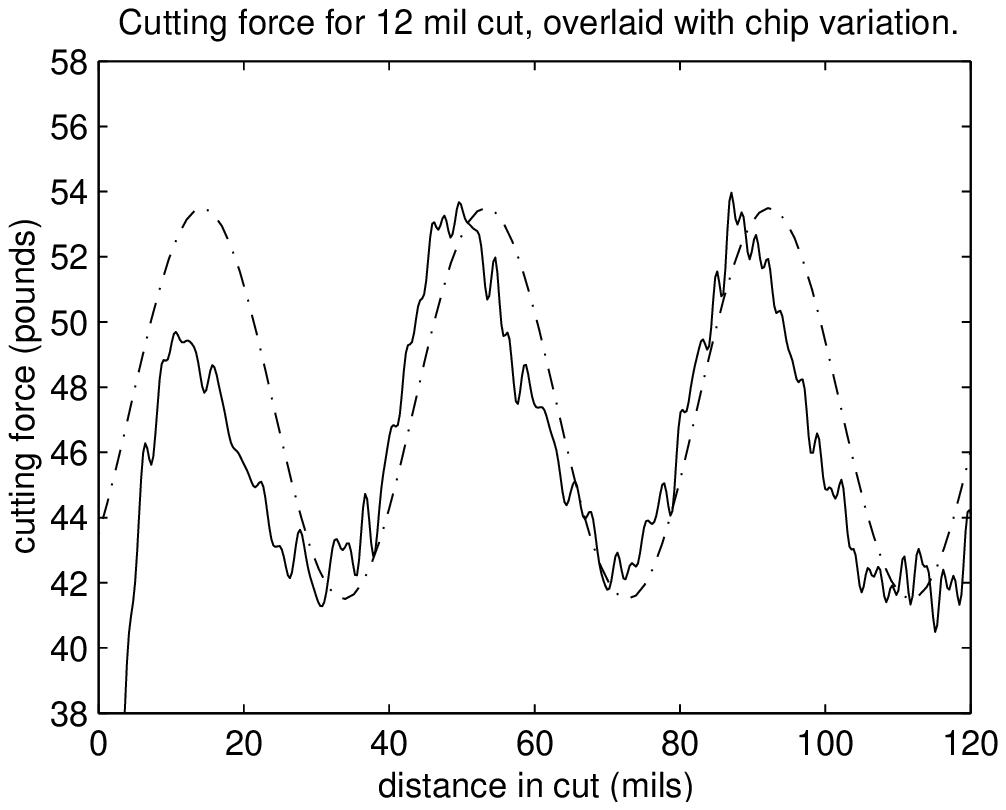}}
\caption{Cutting force vs. distance into cut, overlaid with chip thickness (dashed line).}
\label{cutforce}
\end{figure}

\begin{figure}
\centerline{\epsffile{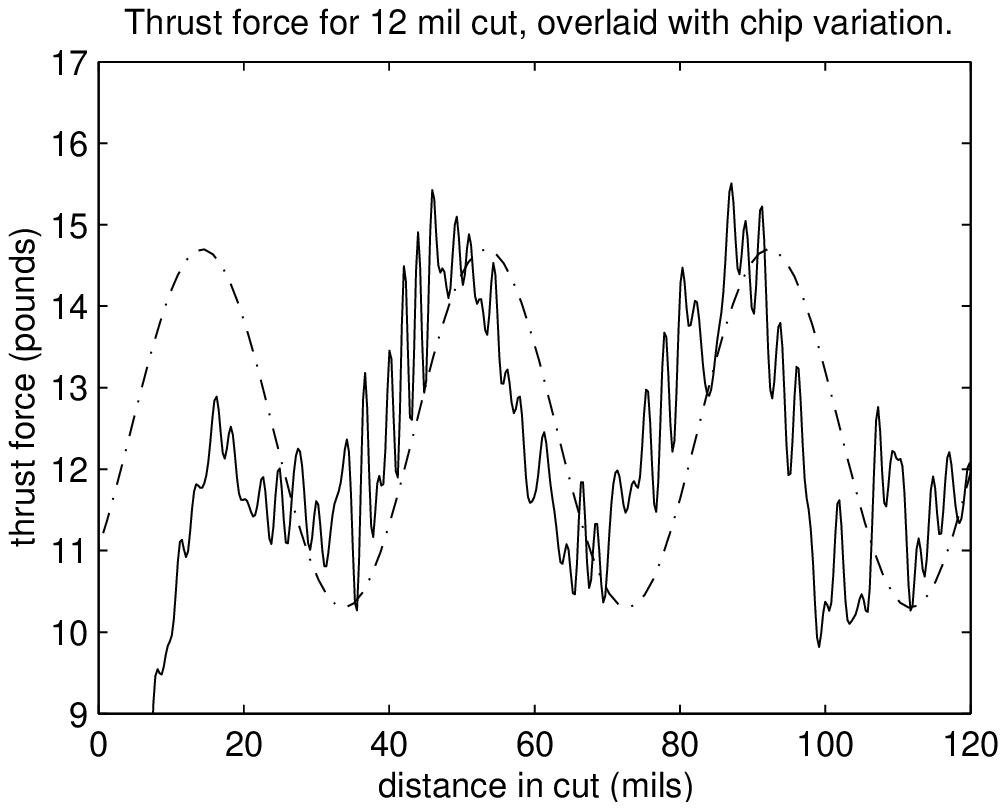}}
\caption{Thrust force vs. distance into cut, overlaid with chip thickness (dashed line).}
\label{thrust}
\end{figure}

Our first observation is that the measured forces in $x$ (cutting) and $y$ (thrust), while 
oscillating at the same frequency as the chip thickness,  lead it in phase.
To illustrate this we plot the thrust and cutting force through the cut, overlaid
with a sine wave of the same frequency and phase as the chip thickness, see 
figures \ref{cutforce} and \ref{thrust}.
(The amplitude and DC offset of the sine wave were scaled to match the force
for illustration).

The phase lead makes sense after analyzing individual frames in the simulations.
(See figure \ref{frames}.) The 
chip begins to bunch up and hence offer more resistance, which creates larger
cutting forces prior to the maximum in the chip thickness.  The forces
drop before the tool tip encounters the next minimum of the chip thickness,
in reaction to the shearplane snapping down to meet the minimum,
which causes the chip to narrow down and offer less resistance.  

This phase lead of the force over the chip thickness has been observed in experiments carried out
in Dr. Philip Bayly's lab at Washington University  (for representative
publications from Dr. Bayly's group see \cite{BAYLY1, BAYLY2, BAYLY3}).   A tube cutting
experiment with a vibrating tool was set up on a lathe, creating an essentially
single degree of freedom system with one dominant mode.  Tests were 
performed to determine the frequency response (FRF) of the cutting system,
and these FRF were used to parameterize a single degree of freedom
model for the cutting process.  This model was developed in part by
Brian Whitehead, who describes in his thesis \cite{WHITEHEAD} an attempt to include friction on
the rake face of the tool by incorporating a weighting function that
models the variation of forces as the chip  slides up the rake face of
the tool, after St{\'e}p{\'a}n \cite{Stepan}.  One of the parameters of the model
is the contact length, $L$, between the tool relief face and chip.  With positive $L$ the weighting functions,
once incorporated into the frequency domain model, act as  low-pass
filters, and induce a lag in the force behind the uncut chip thickness.
Upon fitting this model with the tube cutting experiment FRF, they determined that the best
fit came with using a {\it negative} $L$ value, essentially making the
forces on the tool {\it anticipate} the chip thickness.  This does not
make sense if $L$ is considered to be the contact length on the rake face, but with $L$
negative the force is being integrated over a region out in front of
the tool, which induces the phase lead of the force over the uncut
chip thickness.  We believe that this is a manifestation of the shearplane
oscillation effect we see in the Third Wave cutting simulations.

\begin{figure}
\centerline{\epsffile{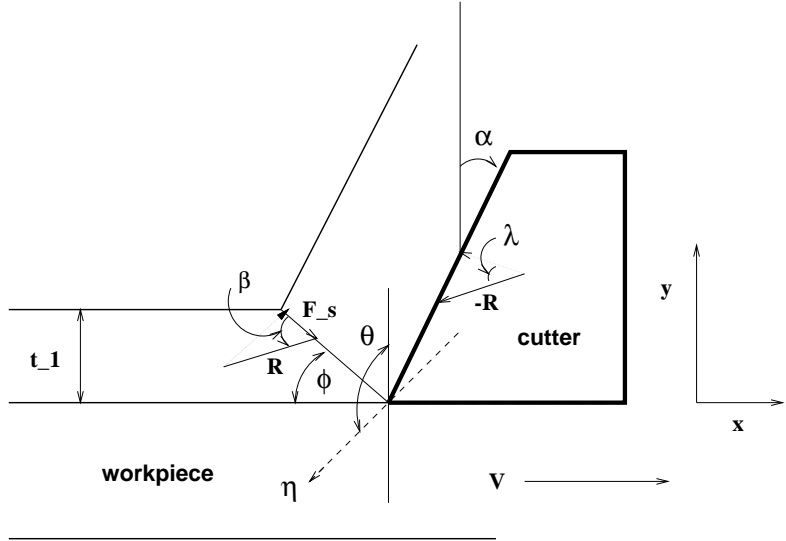}}
\caption{Diagram of angles and forces for orthogonal cutting.}
\label{cut}
\end{figure}

It is plausible that while the forces lead the chip thickness, they may be
in phase with the shearplane length, following a model popularized by Ernst and Merchant
\cite{ERNST},
that assumes that the forces will be proportional to the shear plane length.
The Merchant model predicts that specifically the shear plane force $(F_s$)
is linearly dependent on
the shear plane area, so $F_s = \tau w l$, where $\tau$ is the shear stress
involved in cutting the material, $w$ is the cutting width, and $l$ is the length
of the shear plane, which is oriented at the angle $\phi$ (refer to figure \ref{cut}).
Also from the figure note that $F_s = ||{\bf R}|| \cos \beta$, where ${\bf R}$ is the
total force on the tool.
To determine $\beta$ we rely on the traditional decomposition of ${\bf R}$
into vertical and horizontal components, $F_y$ and $F_x$, the radial or thrust,  
and the tangential or cutting, forces
respectively. 
If the angle that ${\bf R}$ makes with the horizontal is given by $\psi=\arctan(\frac{F_y}{F_x})$,  
then $\beta = \phi+\psi$.
The AdvantEdge program computes $F_x$ and  
$F_y$, which also gives us a measure of the total cutting force as
the tool moves through the cut, since $||{\bf R}|| = \sqrt{F_x^2+F_y^2}$, as well as 
the angle $\psi$.
We can measure the shear plane length $l$, and the angle $\phi$, for each step in the run, and
since $F_s = ||{\bf R}|| \cos (\phi+\psi)$, we can test the Merchant prediction:
\begin{equation}
F_s = ||{\bf R}|| \cos (\phi+\psi) = \tau w l.
 \label{fitfs}
\end{equation}

\begin{figure}
\centerline{\epsffile{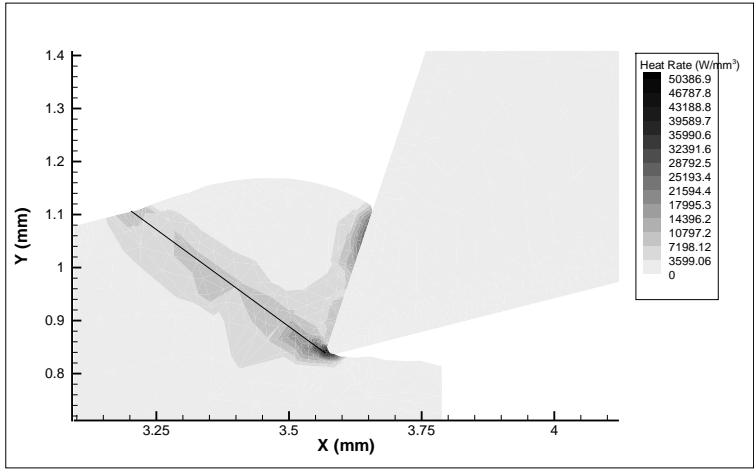}}
\caption{How the shearplane angle and length is computed: the angle of the line 
shown here with the horizontal and its length are measured directly from the frames of
the Third Wave animation.}
\label{shearplaneex}
\end{figure}

\begin{figure}
\centerline{\epsffile{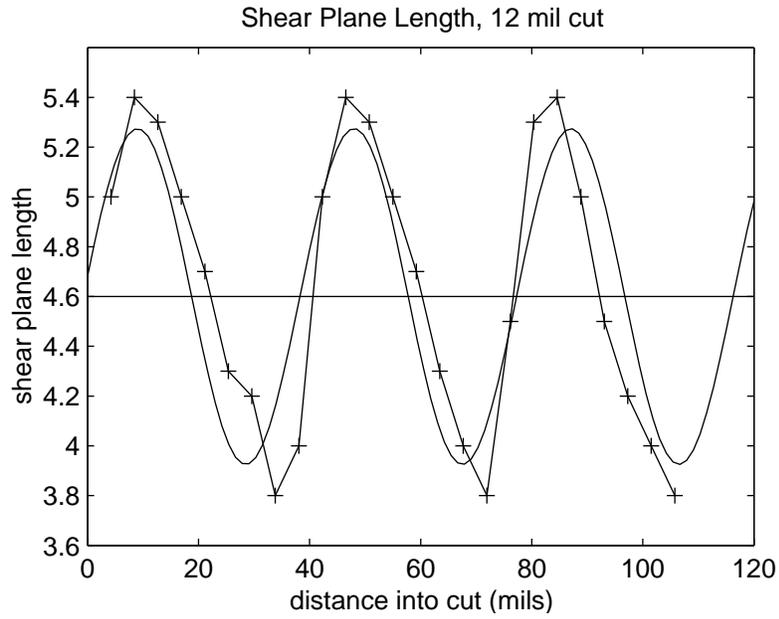}}
\caption{Shear plane length measured from experimental snapshots, overlaid with fitted sine wave.}
\label{fitlength}
\end{figure}

The shearplane length was measured directly from the 
animation frames, where visualizations of the heat production are best for identifying
the shearplane region.  This is subject to obvious error through the identification
of a line that could be called the shearplane.  We chose to measure a line that 
started at the tip of the tool and ended at the surface of the workpiece,
parallel with the contours of heat production.
See figure  \ref{shearplaneex} for an example. 
The shear plane length oscillates along with the undulations in the 
surface of the workpiece, though not necessarily with the same phase.
Following this observation, the shearplane length as the tool moves through the 
workpiece was least squares fit to 
a sine wave with the
same wavelength as the wavy surface.
The result of this exercise for the 12 mil feed run is plotted in figure  \ref{fitlength}, 
and  the fitted
expression is

\[
l(x)=4.6+1.1(.151 \cos(2 \pi x/40) + 0.647 \sin( 2 \pi x/40)).
\]
The units on shearplane length are arbitrary, and vary with the three runs examined
in this section.  The important feature is the phase of the oscillation.

\begin{figure}
\centerline{\epsffile{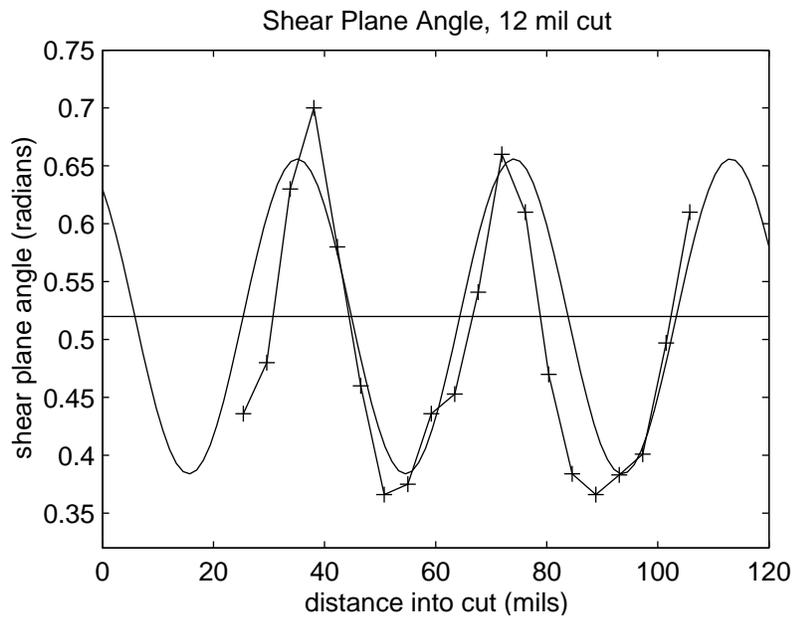}}
\caption{Shear plane angle measured from experimental snapshots, overlaid with fitted sine wave.}
\label{fitangle}
\end{figure}

\begin{figure}
\centerline{\epsffile{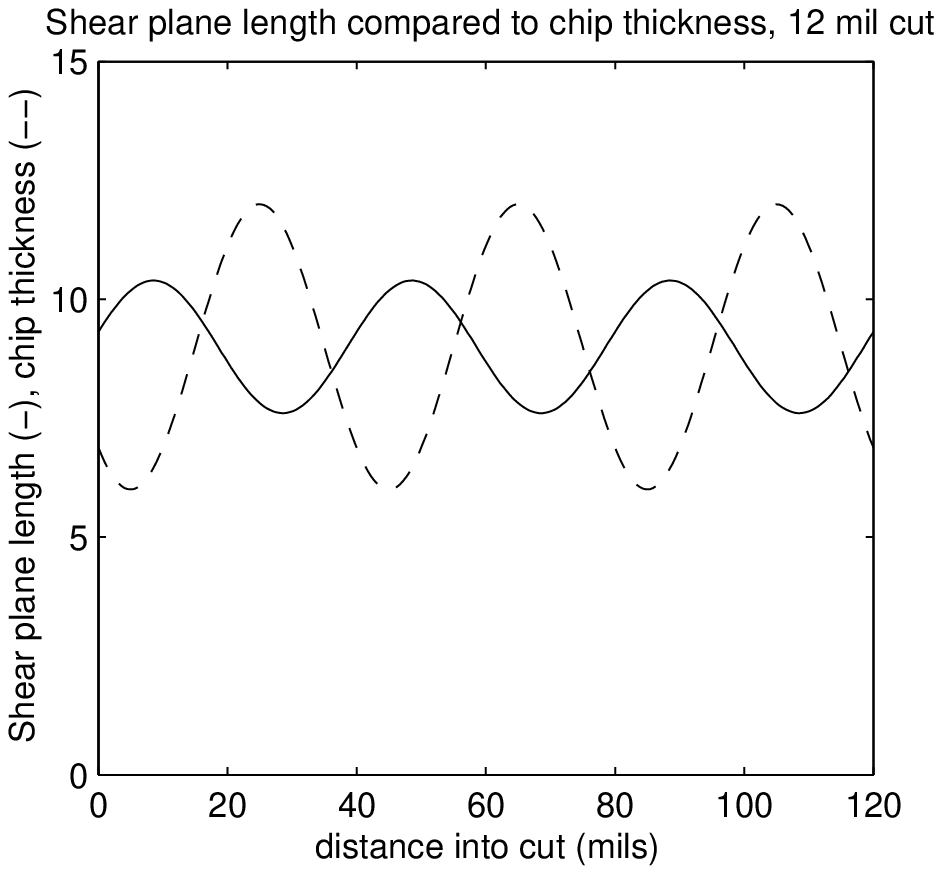}}
\caption{$l(x)$ compared to chip thickness through the cut for 12 mil cut. 
The magnitude of the chip thickness is in mils, the shearplane length has
been scaled and shifted vertically for comparison.
The critical feature 
is the phase shift: the shearplane length leads the chip thickness.}
\label{comparewaves}
\end{figure}

The shearplane angle can be measured and fit in a similar way, the result is presented
in figure  \ref{fitangle}, and the fitted expression is

\[
\phi(x)=0.49+0.1 \cos(2 \pi x/40) - 0.09 \sin (2 \pi x/40).
\]
Similar fits were performed for the 6 and 9 mil cuts, and the units are radians.
We also observe that the shearplane length oscillation leads the chip thickness
oscillation by a noticeable amount, see figure  \ref{comparewaves}.  

\begin{figure}
\centerline{\epsffile{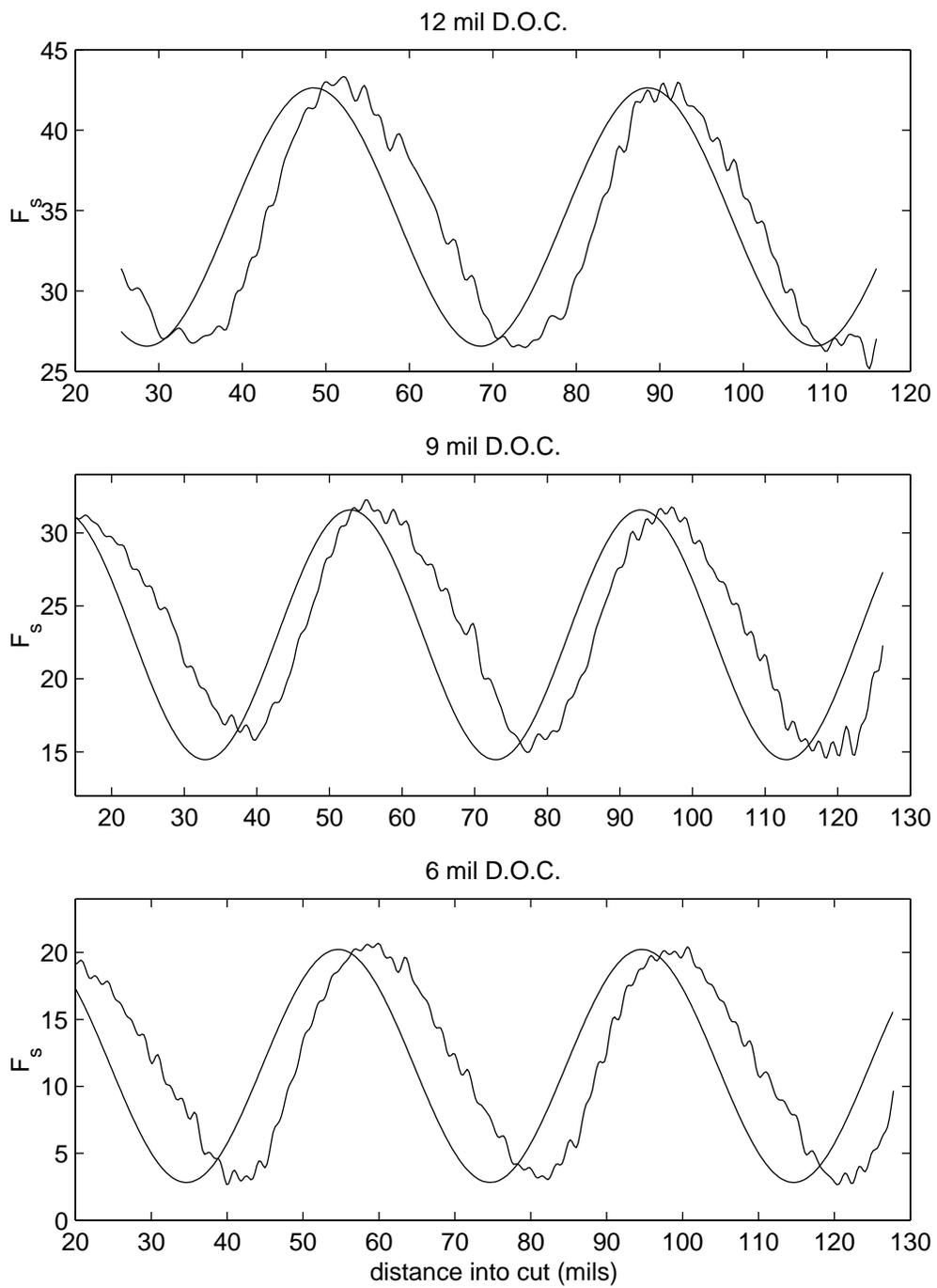}}
\caption{$F_s$ compared with fit to shearplane length for 12 mil, 9 mil and 6 mil
depth of cut (D.O.C.) runs, the length is the smooth curve.}
\label{fsfit}
\end{figure}

We compute $||\bf{R}||=\sqrt{F_x^2+F_y^2}$ and $\beta=\phi+\arctan(\frac{F_y}{F_x})$ 
for each of the three runs as a 
function of distance into the cut ($x$).
Expression (\ref{fitfs}) implies that $l(x) \sim ||{\bf R(x)}|| \cos \beta(x)$, 
so we conducted a least squares fit to $||\bf{R}|| \cos \beta(x)$ with 
the expression $c_1 + c_2 l(x)$,
the result  is presented in figure \ref{fsfit}.  
The shear plane force is seen to lag  the shear plane length, which fits
with the observation that while the shearplane force leads the chip thickness,
the shearplane length leads the chip thickness by an even greater amount,
leading to the lag in shearplane force with respect to the shearplane length.
Also note that we fit the force $F_s$
not with $c_1 l(x)$, as is specified by Ernst and Merchant, instead it is a linear fit offset by
a constant. This indicates that there must be a nonlinear region near $l = 0$, a roll-off
of the function so that $F_s = 0$ when $l =0$.    This by itself does not contradict the
Merchant model, which was meant to apply only in steady cutting finite chip thickness
regimes.  However,
the Merchant formula does fail to completely capture the behavior of the forces
in this dynamic situation.

\begin{figure}
\centerline{\epsffile{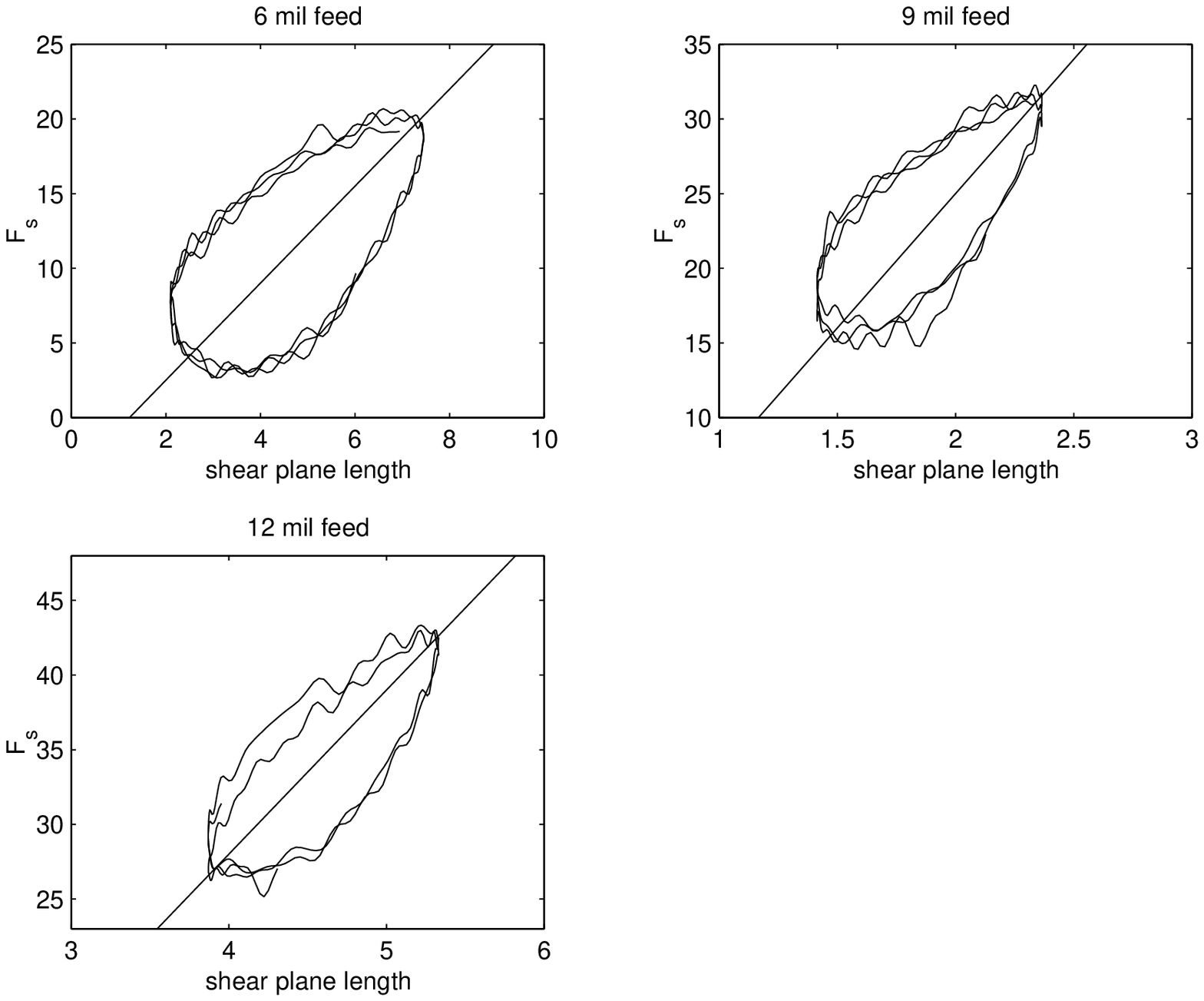}}
\caption{$||\bf{R}(x)|| \cos \beta(x)$ plotted parameterically 
with $x$ against $l$, 6,9 and 12 mil feed.
The shear plane length is in different units in each figure, the measured
shearplane force is in pounds.}
\label{fitline}
\end{figure}

This procedure suggests plotting the data $||{\bf{R}}(x)|| \cos \beta(x)$ parametrically against $l(x)$
to determine if this linear fit is appropriate, see figure \ref{fitline}.  
In this format it is clear that the shearplane force
depends on something in addition to $l$, since it is a multi-valued.
The fitted line is drawn for comparison  for each of the three runs.

\begin{figure}
\centerline{\epsffile{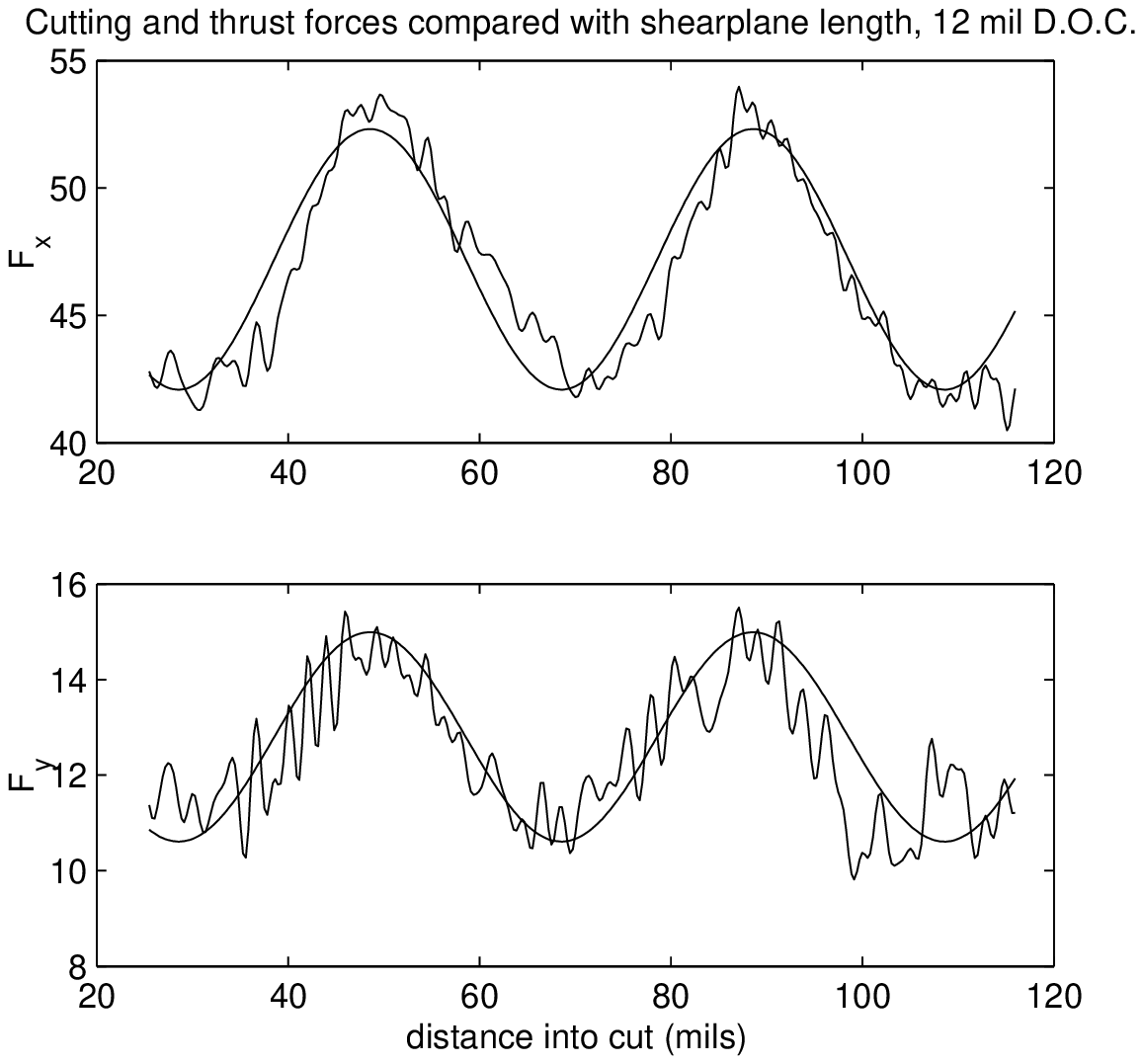}}
\caption{$F_x$ and $F_y$ plotted with $l$ (the smooth curve), 12 mil feed.}
\label{shearcutfit}
\end{figure}

The direct measured forces, $F_x$ and $F_y$ can also be compared with the shearplane
length oscillation during the run.  The result is presented in figure \ref{shearcutfit},
for
the 12 mil chip thickness run.
The cutting force also lags the shearplane length, similar to $F_s$, predictably
since the cutting force is the largest component of the total force.  The thrust
force, however, comes very close to matching the oscillation of the shearplane
length, a fact that could be explored further.

\begin{figure}
\centerline{\epsffile{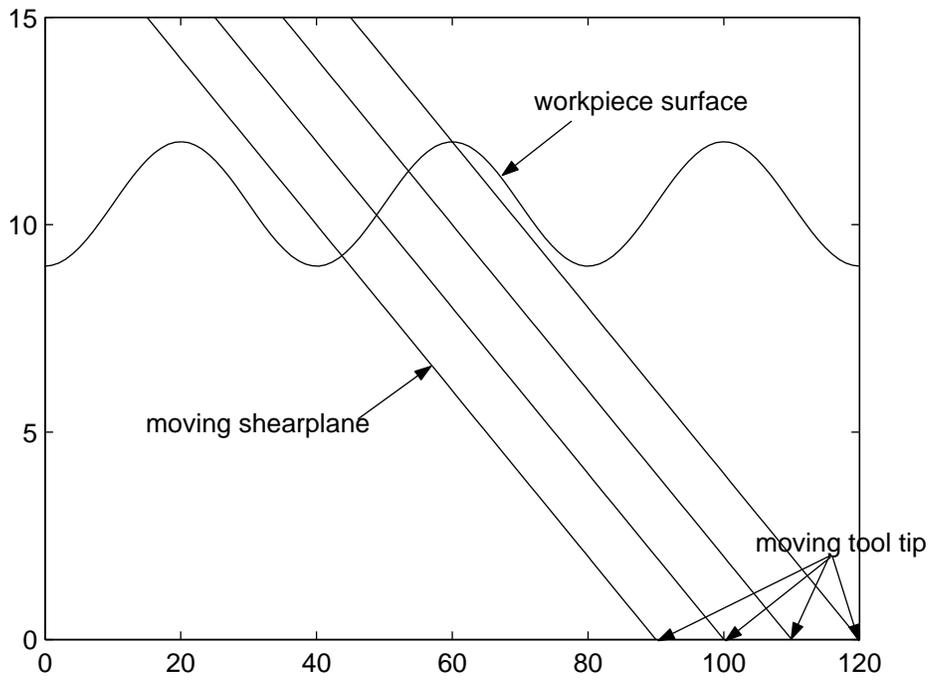}}
\caption{How the shear plane length might change through a wavy surface cut.}
\label{slidingline}
\end{figure}

\begin{figure}
\centerline{\epsffile{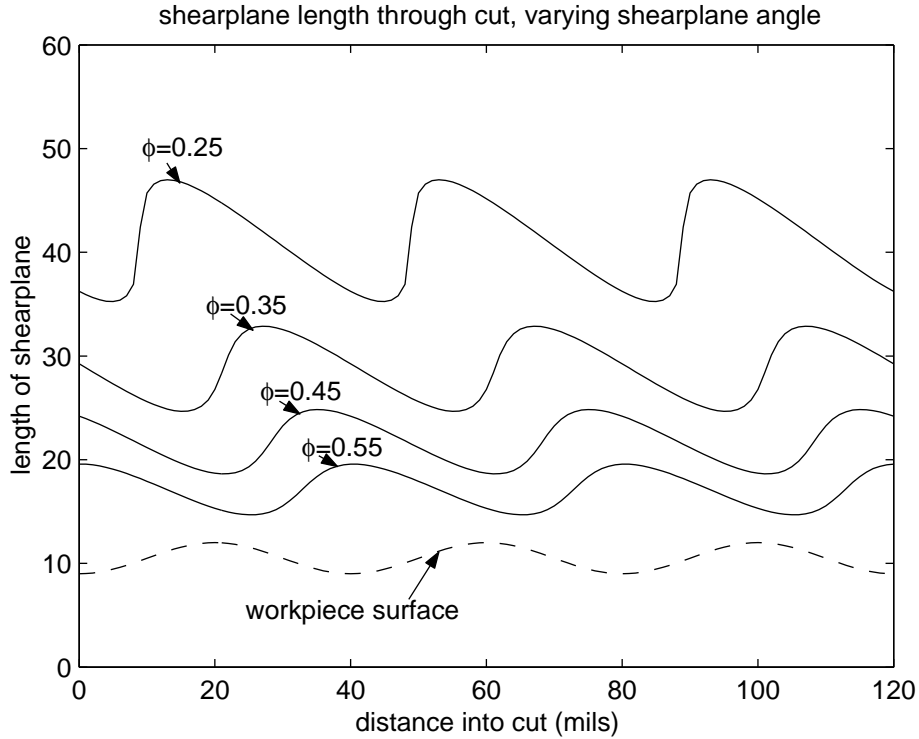}}
\caption{Plotting the shear plane length for varying shear plane angle ($\phi$), through a wavy surface cut.
The function plotted with a dashed line is the workpiece surface itself, for comparison with
the chip thickness.}
\label{slidinglinelength}
\end{figure}

To conclude this section, we would like to comment on an idea presented in 
Wellbourne and Smith \cite{Wellbourne}. They computed an expression for the cutting force involved
in cutting a wavy surface by assuming that the force was proportional to the
shear plane length also.  However, they were limited to estimating the shearplane
length by computing the length of a line, oriented at a constant slope, as it
 intercepts a sine wave (see figure \ref{slidingline}).
The length can be found
by solving a transcendental equation
 (the intersection of the moving line and a sine wave) numerically for
varying shear plane angles (figure \ref{slidinglinelength}).
From this you can see that the length leads the oscillation of the chip load,
how much depending on the shear plane angle.  The oscillation also becomes
more triangular as the angle is decreased, it will have a discontinuity when the multiple roots to the transcendental
equation appear, that is, for small enough shearplane angles.  

These effects certainly play into the form of the forces determined with experimentally measured
shear plane lengths.  But there are several confounding factors:  the shear plane angle is not
a constant, and more importantly, the surface itself is deformed during a cut, and the
simple intersection of a line and a sine wave does not give the correct shear plane length,
as we saw in the animations, see figure \ref{frames}.

\section{Analyzing Crushing Forces}
As described in the Introduction, the new Third Wave module that allows the 
tool to oscillate either horizontally or vertically during a cut gave us
an opportunity to explore the forces involved when the relief face of the cutting
tool impacts the workpiece.  
  In our simulations to investigate relief face crushing the cutting
tool was vibrated sinusoidally in the vertical direction with a 3 mil amplitude 
and varying frequencies.
Two simulations were performed to calculate the crushing force for each set
of cutting parameters,  identical in all
ways except for the relief angle of the tool.  In one simulation,
the relief angle was small enough that the entire relief face came
into contact with the workpiece, causing the crushing of the workpiece
at the relief face of the tool.  In the other simulation,
the relief angle was made large enough so that there was no
contact between the relief face of the tool and the work piece.

\begin{figure}
\centerline{\epsffile{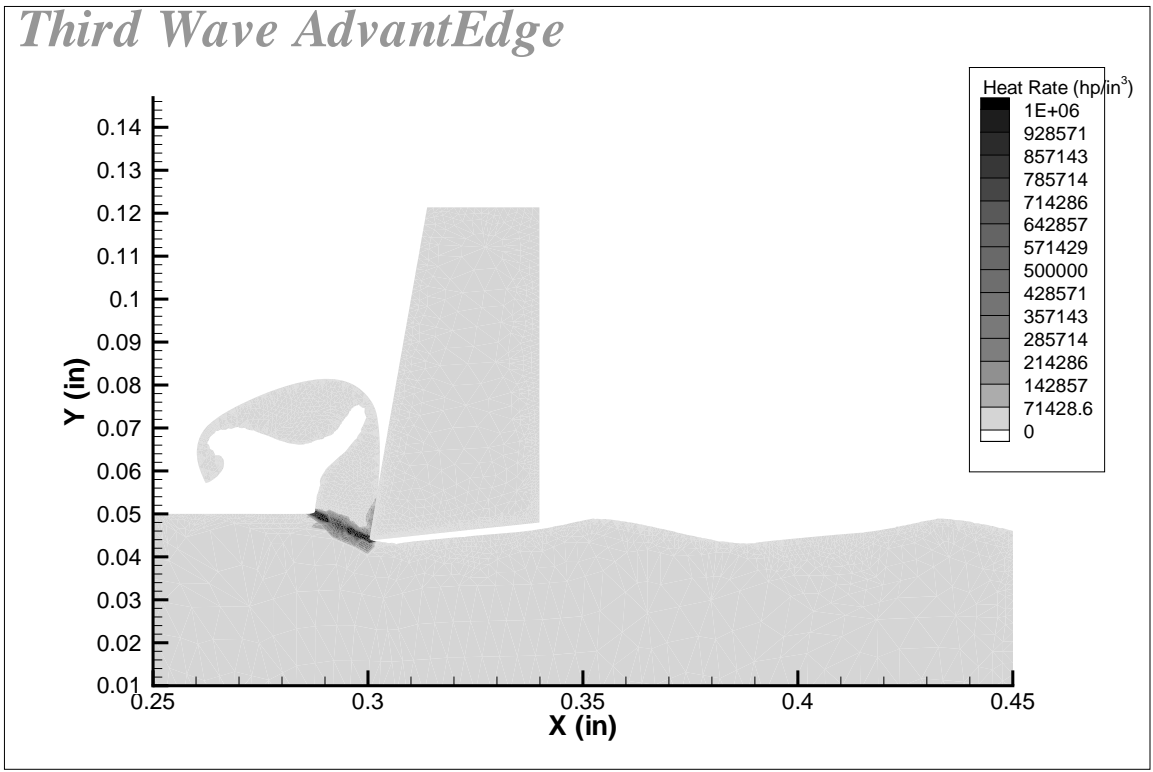}}
\caption{Third Wave run (AL7050) with relief angle of 6 degrees.}
\label{crushrun}
\end{figure}

\begin{figure}
\centerline{\epsffile{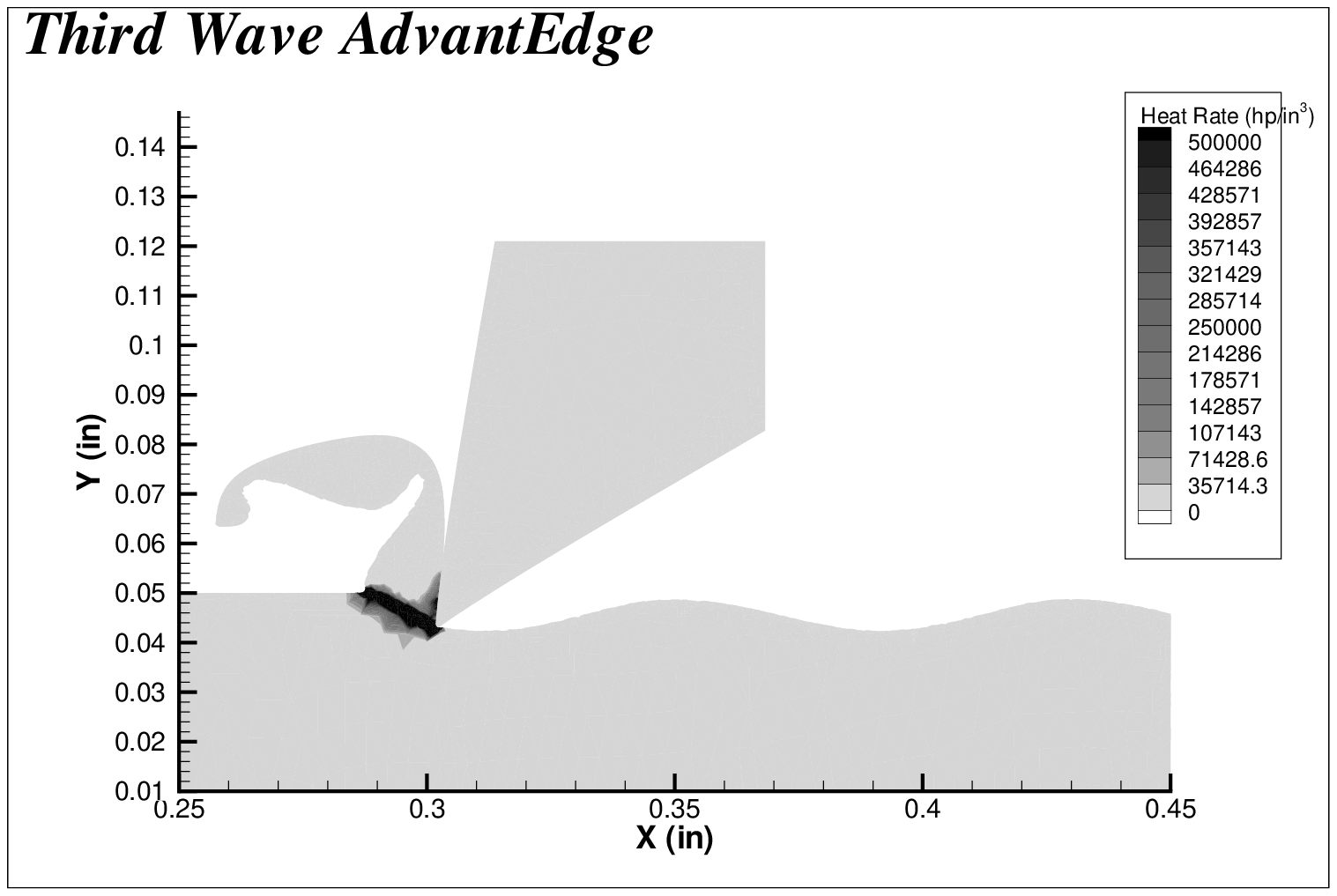}}
\caption{Third Wave run (AL7050) with relief angle of 25 degrees.}
\label{nocrushrun}
\end{figure}

\begin{figure}
\centerline{\epsffile{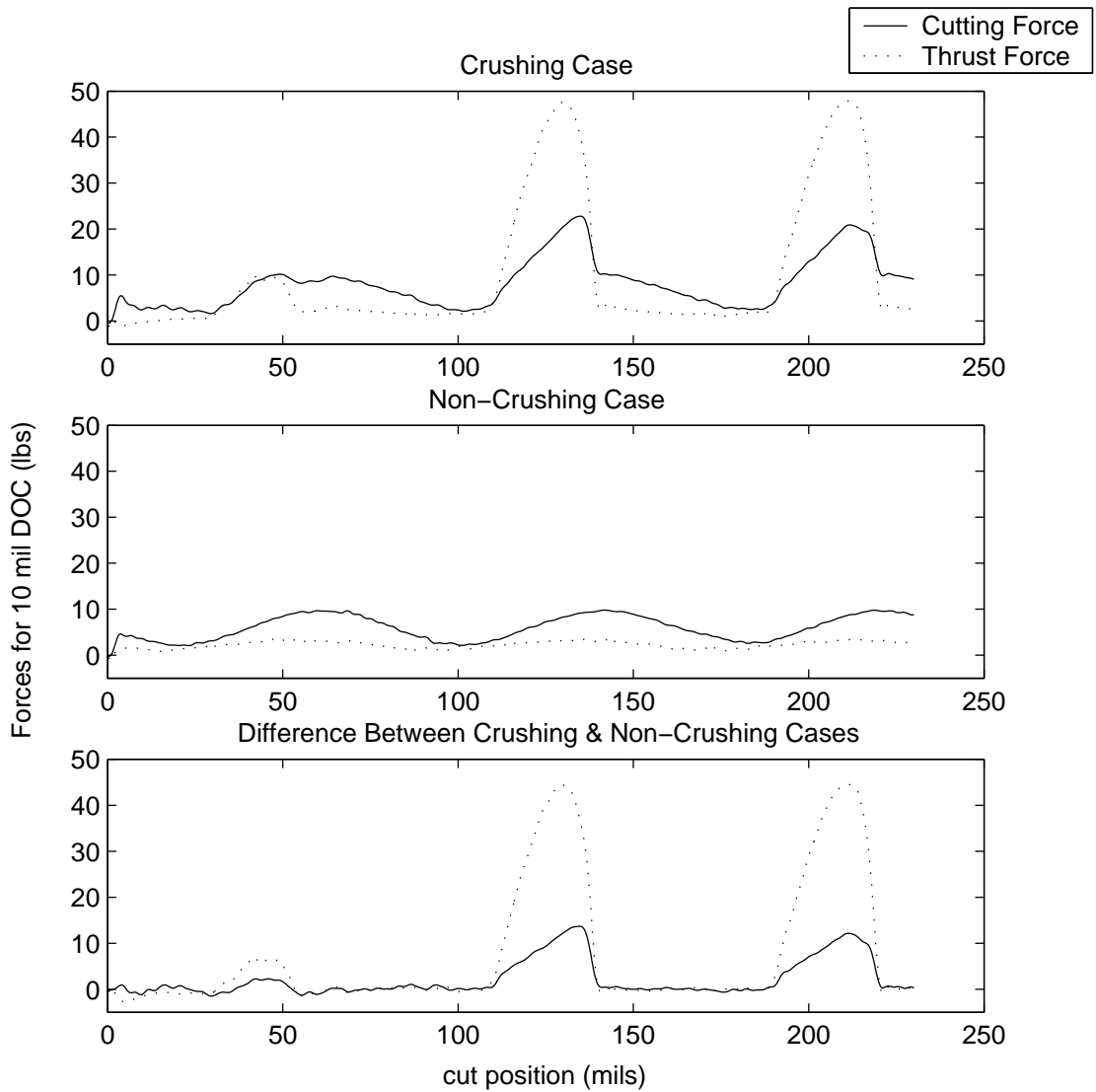}}
\caption{Computing the crushing forces from two Third Wave runs.}
\label{crushnocrush}
\end{figure}

To illustrate, a snapshot of a crushing run is shown in figure \ref{crushrun},
where the tool has come into contact with the workpiece during the downstroke of the
tool and left a flattened region in its otherwise sinusoidal path.  This should be
compared with a run with the relief angle drawn up from 6 degrees to 25 degrees,  
figure \ref{nocrushrun}, where the tool tip leaves a sinusoidal workpiece surface in its
wake.  Assuming the forces involved in material removal and crushing combine linearly,
the crushing force can be resolved by subtracting the forces from two such runs.
The exercise is illustrated in figure \ref{crushnocrush}, where the material 
being cut is AL7050, with a tool with rake angle of 10 degrees and cutting edge radius of 0.5 mils.
In the two runs the wavelength of the oscillation is 80 mils, corresponding to
a cutting speed of 2680 SFM and vibration frequency of 6.7 kHz.  The nominal feed
is 1 mil, and vibration amplitude is 3 mil.  
The spatial wavelength $\lambda$ of the undulating cut is equal to
the ratio of the vibration frequency $\nu$ to the cutting speed $V$, that is
$\lambda=\frac{V}{\nu}$.  
We focused our investigations on the vertical or thrust force in what follows.

We found that the relief edge length, relative to the wavelength of the 
oscillation, was critical in determining the behavior of the crushing force in these runs.
Therefore we ran simulations with tool relief lengths of 10 mil, 30 mil and 100 mil,
which were much smaller, somewhat smaller, and much larger than the range
of wavelengths tested.  The results are summarized below.
 
\subsection{Crushing force simulations}
\begin{figure}
\centerline{\epsffile{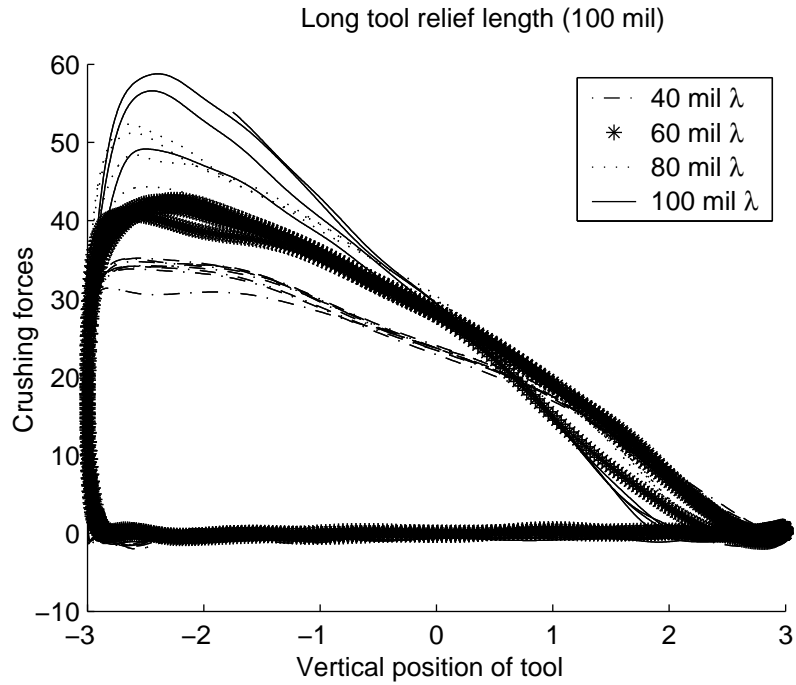}}
\caption{Thrust crushing force vs. vertical position of tool, long tool 
relief length, multiple wavelengths.}
\label{cforcelong}
\end{figure}

\begin{figure}
\centerline{\epsffile{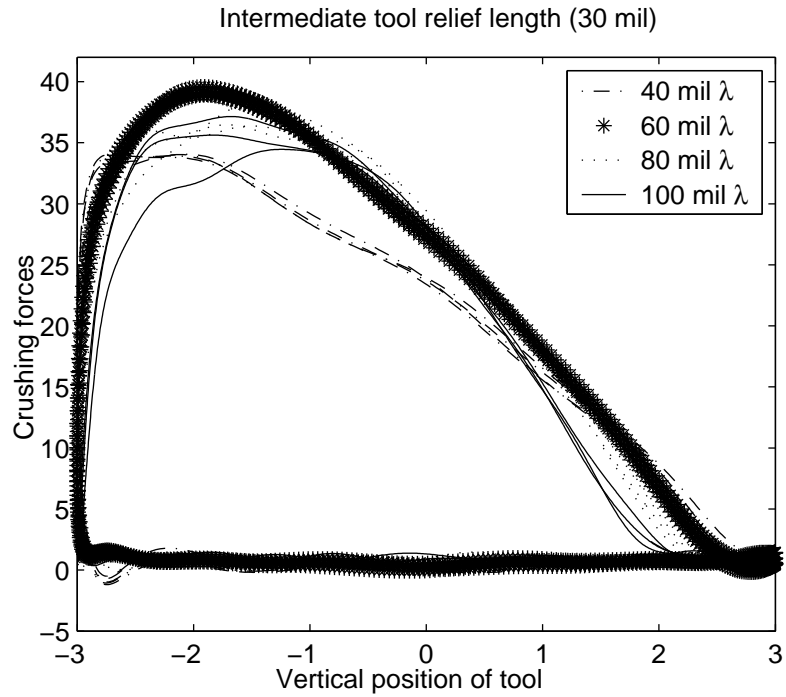}}
\caption{Thrust crushing force vs. vertical position of tool, intermediate tool 
relief length, multiple wavelengths.}
\label{cforcemed}
\end{figure}

\begin{figure}
\centerline{\epsffile{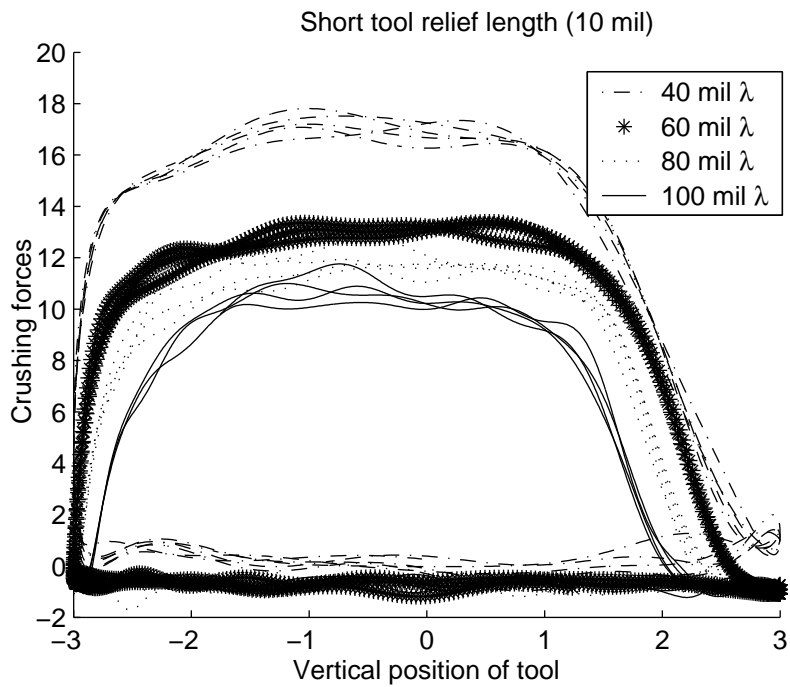}}
\caption{Thrust crushing force vs. vertical position of tool, short tool 
relief length, multiple wavelengths.}
\label{cforceshort}
\end{figure}

Once again the AdvantEdge runs were performed on AL 7050 with a carbide tool with
a 10 degree rake angle and cutting edge radius of 0.7874 mils.  The cutting 
speed was chosen to be 2680 SFM, and vibration frequencies near 13.4 KHz, and
amplitudes of 3 mils, typical of an axial-torsional vibration mode for a twist
drill (\cite{STONE}).  
In total we did 24 runs: 12 crushing, 12 no crushing. For each length of tool relief
face (10, 30 and 100 mils) we ran four wavelengths: 40, 60, 80 and 100 mil.
We found plotting the crushing force vs. the vertical amplitude of the tool
during a run a useful way of visualizing the data.  In  figures 
\ref{cforcelong}-\ref{cforceshort} these results are presented for
each tool relief length, with the varying wavelength runs plotted 
together for comparison.  The general form of each plot is the same,
during the downstroke of the tool, as expected, the crushing force
increases up the angled portion of the loop, i.e. the loops are
traversed counterclockwise with increasing time.  Near the end
of the downstroke the force begins to decrease, and drops to zero
abruptly as the tool begins its trip back up to the top of the oscillation.
During this phase the force is zero and the portion of the loop 
along the displacement axis is traversed. Immediately after  the 
top of the oscillation is reached and the downstroke begins,
the force starts to increase  again, and another cycle is initiated.

The three sets show a qualitative
difference in shape, and a quantitative difference in maximum force
achieved relative to wavelength of the oscillation.  The long relief
length tool demonstrates the most easily understood force loop: the
force increases approximately linearly as the tool moves into the material, 
reaching a maximum close to the lowest point of the traverse.
In the short relief length tool runs the force increases approximately
linearly until about one third of the way down, when it 
flattens out.  The short tool relief length is saturated at  this point:
is has come into contact with as much material as it can, it cannot 
impact more length as it moves down.   The intermediate length tool runs 
have a shorter flat region that depends on wavelength, with longer wavelengths
showing some flattening.


\begin{figure}
\centerline{\epsffile{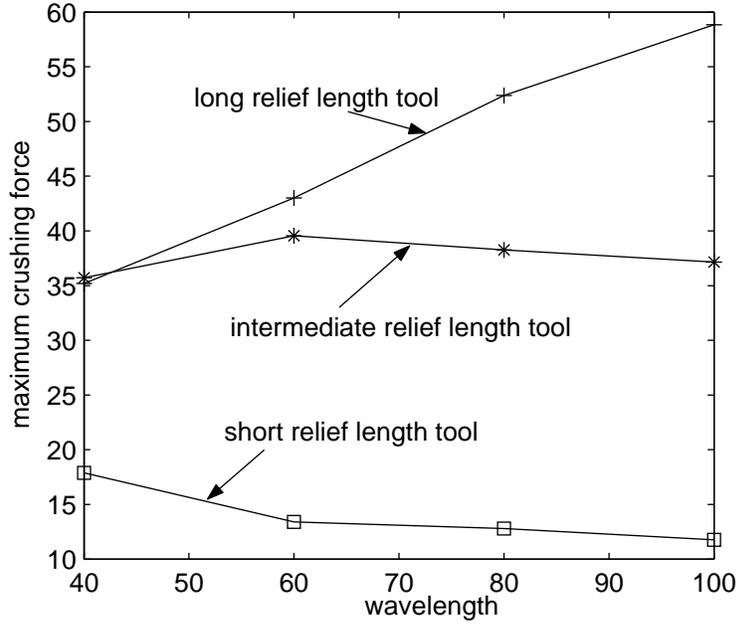}}
\caption{Maximum crushing force vs. wavelength of oscillation.} 
\label{wldep}
\end{figure}

The quantitative difference between the three sets of runs concerns the
dependence of the maximum force value vs. wavelength of oscillation.  In Whitehead \cite{WHITEHEAD},
a model for process damping due to rubbing on the relief face is proposed that depends inversely on the
wavelength of the vibration, based on a suggestion by Tobias \cite{Tobias}. 
This translates to a term in his linear model that is directly proportional 
vibration velocity (which depends directly on vibration frequency) and 
inversely proportional to cutting speed.
In figure \ref{wldep} we plot the maximum force
for each run vs. wavelength in the three cases.  Though the data is too
sparse to fit any detailed model, it is clear that the force is inversely proportional
to the wavelength only in the short relief length runs, the long relief
length runs demonstrate a direct dependence on wavelength, and in the
intermediate length runs the maximum force appears to be indifferent
to wavelength.  The dependence of the damping on cutting speed and vibration frequency thus
will depend on the length of the relief face relative to the vibration wavelength,
and given that chatter vibrations are mainly high frequency, the long tool
data will be the most relevant for estimating chatter process damping.

\subsection{Crushing force modeling}
\begin{figure}
\centerline{\epsffile{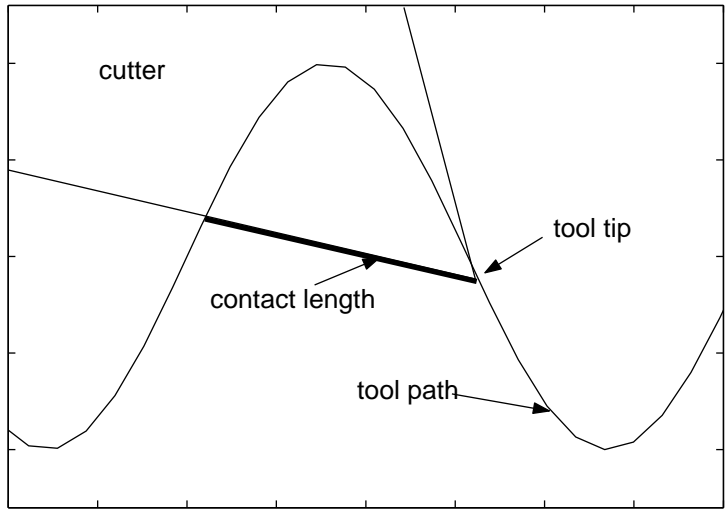}}
\caption{Illustration of contact length during crushing, long tool relief length.}
\label{intersection}
\end{figure}

The magnitude of the crushing force will most certainly depend on the amount of
contact between the relief face of the tool and the material.  We can
analytically model the variation of the contact during a crushing thrust 
of the tool by determining the contact length, as it depends on frequency of
the oscillation, height of oscillation and the cutting speed.
The contact length at each point along the cut can be
calculated from the intersection of a tool edge with the
work piece (see figure \ref{intersection}).
The sinusoid in the figure is the path that the tool takes during
an oscillation, the chip is not shown in the diagram.
A small program in MATLAB calculated this intersection length in
the figures that follow.

\begin{figure}
\centerline{\epsffile{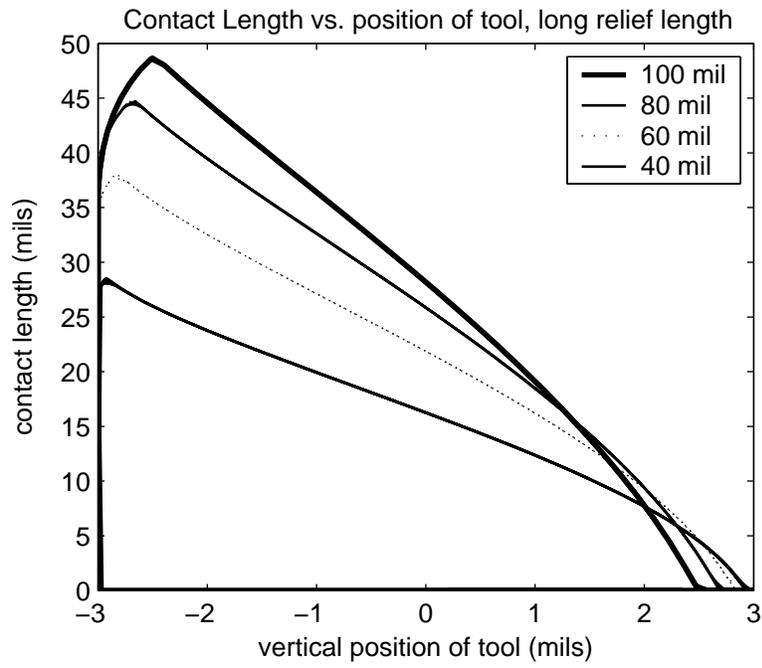}}
\caption{Contact length during tool oscillation, long tool relief length (100 mils).}
\label{CLL}
\end{figure}

The contact length during an oscillating cut is plotted vs. vertical 
position of the tool tip, see figure \ref{CLL}. 
This representation can be directly compared to the force vs.
vertical position plot for the 3rd Wave run with a 100 mil relief length tool,
figure \ref{cforcelong}.   The contact length and the force have
similar dependence on wavelength, both increase with increasing wavelength at the bottom of the downstroke
of the tool (the highest point on the loop).  There is also a ``roll-over" at the top of
the downstroke, where the shorter wavelength force/contact length does not begin to grow as
soon as the longer wavelength case.  This is a consequence of the relative slope of the  
tool path to the relief angle of the tool.
A similar diagram for the short relief length (10 mil) tool is shown in figure \ref{CLS}.
The contact and the crushing forces in this case saturate when the tool relief length is
fully engaged with the workpiece, but an obvious difference between the two is the 
invariance the maximum contact length shows with respect to wavelength.  The maximum forces, 
however, decrease with increasing wavelength for the short tool.

\begin{figure}
\centerline{\epsffile{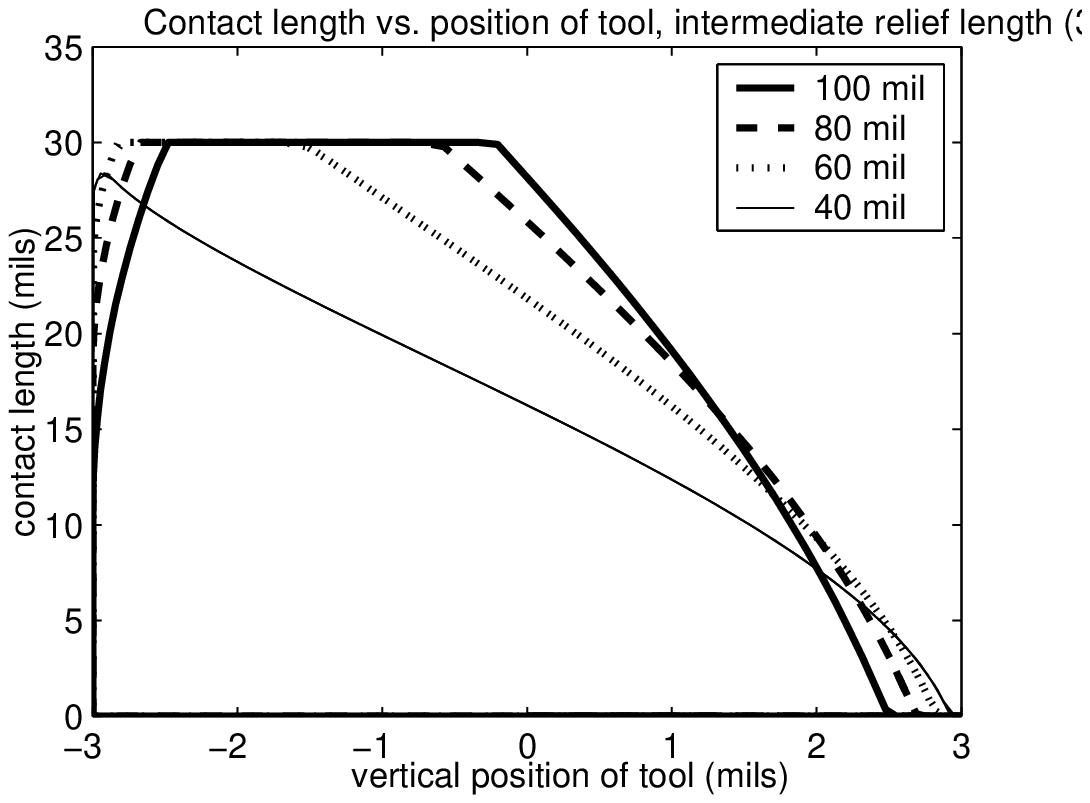}}
\caption{Contact length during tool oscillation, long tool relief length (30 mils).}
\label{CLM}
\end{figure}

\begin{figure}
\centerline{\epsffile{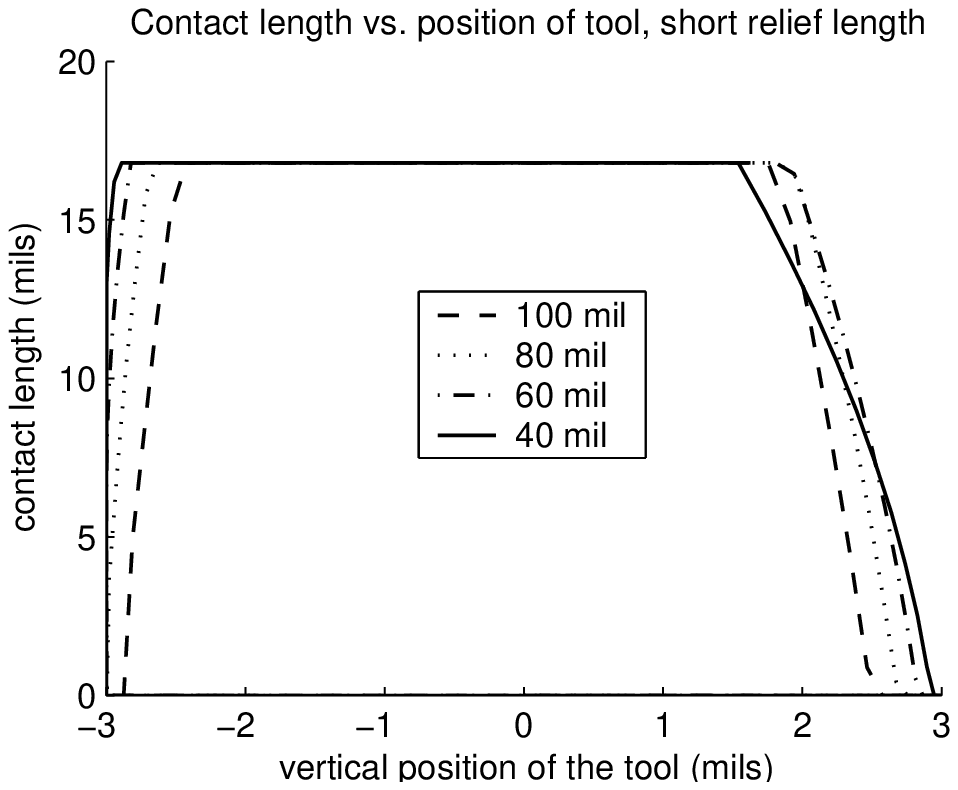}}
\caption{Contact length during tool oscillation, short tool relief length (10 mils).}
\label{CLS}
\end{figure}

With this evidence in mind we can attempt to fit the crushing force data from the experiments
with a model that has a linear dependence on contact length.   
If $X$ is the contact length, $Y=m X + b$ is the crushing force, and $m$ and $b$ are
fitted via least squares.  The results depend on the wavelength of the oscillation
and the relief length of the tool, and are tabulated below.

Long relief length tool  (100 mil): \\
\begin{center}
\begin{tabular}{|l|l|l|} \hline
wavelength (mils) & fitted line & error (lbs)\\
\hline

40  &     Y = 1.102 * X  + 0.397  & 2.34\\

60 &
        Y = 1.132 * X  - 1.539 & 3.75\\
80 &
        Y = 1.184 * X  - 9.698 & 3.88\\
100 &
        Y = 1.305 * X  - 20.493 & 3.74\\
\hline
\end{tabular}
\end{center}

Intermediate relief length tool  (30 mil): \\
\begin{center}
\begin{tabular}{|l|l|l|} \hline
wavelength (mils) & fitted line & error (lbs)\\
\hline
40 &     Y = 1.002 * X  +  3.628 &  3.02\\

60 &
        Y =  1.431 * X  - 6.472 & 2.54\\
80 &
        Y =  0.963 * X  - 7.128 &  2.47\\
100 &
        Y = 0.977 * X  -  7.877 & 2.62 \\
\hline
\end{tabular}
\end{center}

Short relief length tool  (30 mil): \\
\begin{center}
\begin{tabular}{|l|l|l|} \hline
wavelength (mils) & fitted line & error (lbs)\\
\hline
40 &     Y = 0.548 * X  + 1.501 & 0.84 \\

60&
        Y = 0.234 * X  + 1.350 & 0.51\\
80&
        Y = 0.207 * X  + 1.484 & 0.12\\
100&
        Y = 0.167 * X  + 0.957 & 0.33\\
\hline
\end{tabular}

\end{center}

The slope constant in the linear relationship gives the scale 
factor between force (in lbs) and contact length (in mils) and
thus has units lbs/mil of contact length.  This conversion
factor varies somewhat within runs (mean and standard deviation
are $1.18 \pm 0.089, 1.093 \pm 0.226, 0.289 \pm 0.175$, 
for the long, intermediate and short relief length runs, respectively)
The y intercept is determined by nonlinearity in the data at
the top and bottom of the cycle in the data, which causes 
the linear region to move up or down. 

To illustrate these results we plot the scaled contact length vs. vertical position
of the tool along with the crushing force for a comparable run, for the three
different relief lengths and varying oscillation wavelengths (figures 22,23, and 24).
The long tool force loop shows a close fit with contact length, except near the top
of the oscillation when the contact length grows more quickly from zero.
The intermediate tool shows a better fit with data at top of the oscillation, though
the force does not flatten out in the same way as the contact length.
The short tool forces are undercut the contact length, the contact length grows
more quickly at the top of the oscillation, again, and the forces begin to decrease
from the plateau before the contact length comes down.  All these point to 
additional factors coming into play in the prediction of crushing forces.
The assumption of linear dependence on contact length holds primarily for the
mid-region of the downstroke of the tool, excluding the lift-off at the bottom
and turn-around point at the top. 

\begin{figure}
\centerline{\epsffile{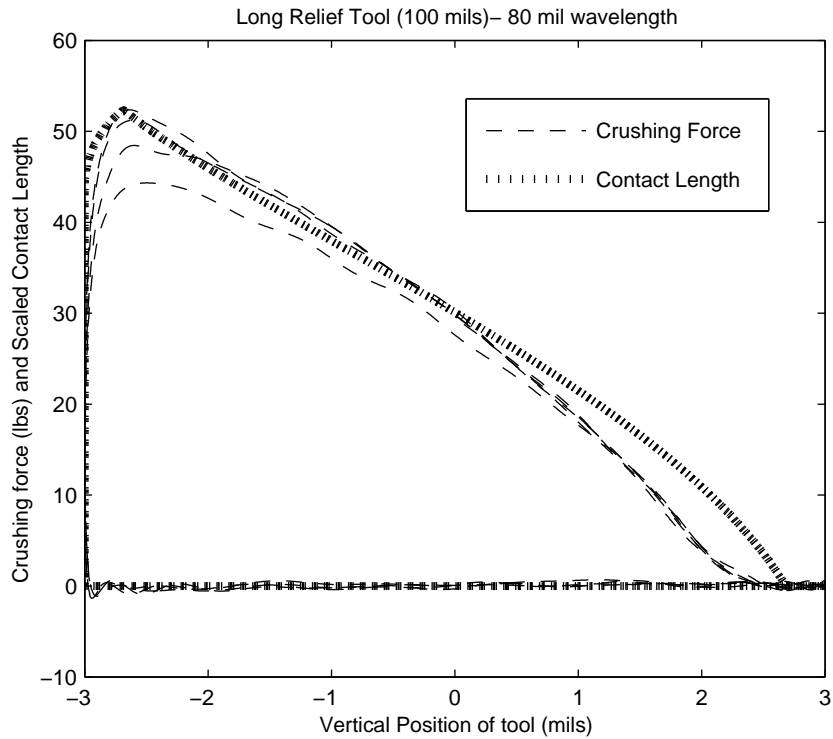}}
\caption{Crushing force and contact length during tool oscillation, long tool relief length,
80 mil wavelength.}
\label{fitlong}
\end{figure}

\begin{figure}
\centerline{\epsffile{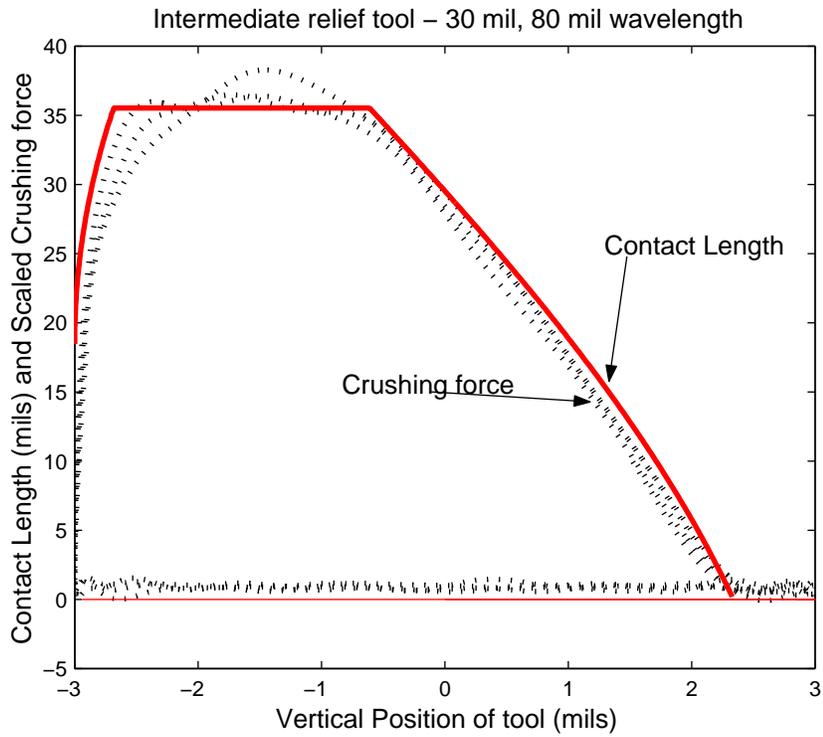}}
\caption{Crushing force and contact length during tool oscillation, intermediate tool relief length.}
\label{fitinter}
\end{figure}

\begin{figure}
\centerline{\epsffile{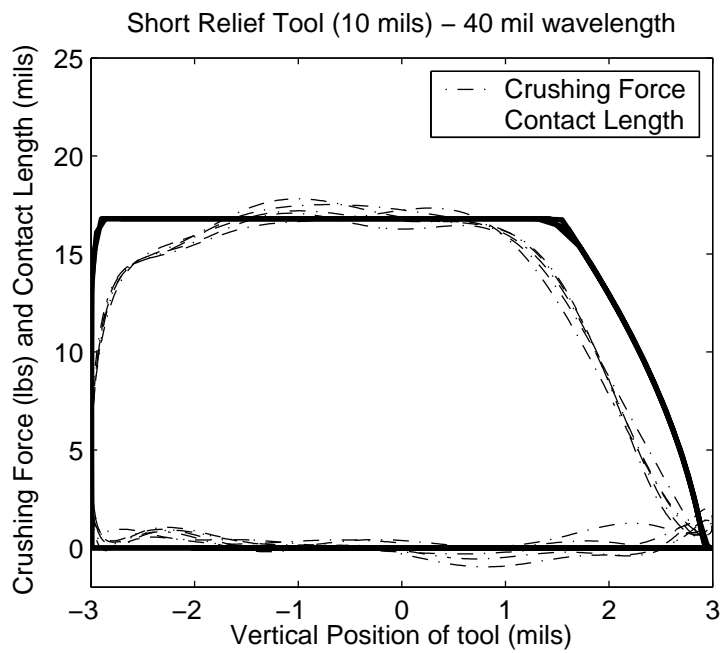}}
\caption{Crushing force and contact length during tool oscillation, short tool relief length.}
\label{fitshort}
\end{figure}

To illustrate these results we plot the scaled contact length vs. vertical position
of the tool along with the crushing force for a comparable run, for the three
different relief lengths and varying oscillation wavelengths (figures 25, 26, and 27).
The long tool force loop shows a close fit with contact length, except near the top
of the oscillation when the contact length grows more quickly from zero.
The intermediate tool shows a better fit with data at top of the oscillation, though
the force does not flatten out in the same way as the contact length.
The short tool forces are undercut the contact length, the contact length grows
more quickly at the top of the oscillation, again, and the forces begin to decrease
from the plateau before the contact length comes down.  All these point to 
additional factors coming into play in the prediction of crushing forces.
The assumption of linear dependence on contact length holds primarily for the
mid-region of the downstroke of the tool, excluding the lift-off at the bottom
and turn-around point at the top.

%In figure \ref{CL} we plot the contact length as a 
%cutting tool with long relief edge moves through a cut at given vibration frequency (6.7 kHz) and cutting speed
%(2780 SFM)
%for three different rake angles.  The vibration amplitude (3.0 mils) and frequency
%were chosen to  reflect observed chatter conditions \cite{Princehouse}.
%{\it it would be helpful to have the oscillation plotted under these graphs...}
%\begin{figure}
%\centerline{\epsffile{figures/contactlength.eps}}
%\caption{contact length vs. position  for a cut with an
%80 mil wavelength, long tool.}
%\label{CL}
%\end{figure}

\section{Discussion and Conclusions}
Our investigations with Third Wave simulations were designed to isolate
different effects present during regenerative chatter vibrations in metal cutting:
the vibration of the tool and the oscillatory variation in the chip thickness.
The tool vibration feature of the package was used to investigate process
damping due to crushing of the workpiece material on the relief face of the
tool, and the custom workpiece feature allowed us to set up a sinusoidally
varying chip-load.  Simulations with the later demonstrated an oscillation
of the shear plane as the tool progressed, and we found that the cutting, thrust
and shear plane forces all were in advance of the chip thickness. The shear
plane force, proposed to be directly dependent on shear plane length by
Ernst and Merchant \cite{ERNST}, was found to lead the shear plane length
as well, as the forces anticipated the decrease in chip load.  This anticipation
is due primarily to the deformation of the chip as it encounters the tool,
not surprizingly the process is more complex than predicted during dynamic cutting.

An important cause of process damping is the interaction of the
relief face of the cutter with the workpiece material.  Here we
studied the thrust forces involved in such damping by running
an oscillating tool in AdvantEdge for a variety of oscillation
wavelengths and tool relief lengths.  We found that the wavelength
dependence of the force was determined by the length of the
relief face of the cutter relative to the wavelength of
the oscillation: for long  relative relief length the maximum force 
increased with increasing wavelength, for short relative relief length it
decreased, and for intermediate relative relief length the dependence
was indeterminent.   This expands the intuition of Whitehead et al., \cite{WHITEHEAD}
who used a process damping model that depended inversely on 
wavelength, as it would for a short tool.  Since the wavelength
of the oscillation equals the cutting speed divided by the oscillation
frequency, this means the damping is proportional to the cutting
speed for short tools, and inversely proportional to the cutting
speed for long tools.  This, however, assumes that the frequency
of oscillation is independent of cutting speed, which is most likely
not the case, given the form of the linear stability diagrams for 
regenerative chatter (for an example related to this study see \cite{STONE}).

We also found that the force depended linearly on contact length 
between the relief face of the tool and the workpiece material 
during the downward stroke of the tool.  Other factors come into play
during the initial  moments of the downward stroke, and through
the ``lift-off" of the tool from the trough of the oscillation.
This suggests that this form of process damping can be explained
purely geometrically,  where the relevant factors are
oscillation amplitude and relative size of relief length to wavelength.
Incorporating this into a model that would determine stability of
the steady cutting state to oscillations would not be straight-forward,
however, since the damping itself depends on the wavelength of the
instability induced (typically the wavelength is determined by
the physical parameters of the system, not a priori).

In summary, this research highlights the importance of dynamic
and nonlinear effects on instabilities in metal cutting, even in
the case of orthogonal cutting.   Forces that act in anticipation
of chipload increases and process damping that depends in a complicated
way on vibration frequency are examples of effects that could
be studied further, both in the laboratory and with large scale
and predictive computer modeling.

\section*{Acknowledgements}
This work was supported in part by the National Science Foundation
(DMS-0104818) and the Mathematics and Computing Technology Group,
The Boeing Company, Bellevue, WA.


\begin{thebibliography}{}

\bibitem{WHITEHEAD} Whitehead, B., The Effect of Process Damping on Stability
and Hole Form in Drilling,  Master of Science Thesis, Sevier Institute of Technology,
Washington University, St. Louis, (2001).

\bibitem{BAYLY1}  Bayly, P.V., S. A. Metzler, A.J. Schaut, and K.A. Young,
Theory of torsional chatter in twist drills: model, stability analysis
and comparison to test, ASME J. Man. Sci. \& Eng., {\bf 123} (4), 552-561 (2001).

\bibitem{BAYLY2} Bayly, P.V., K.A. Young and J.E. Halley, Analysis of
tool oscillation and hole roundness error in a quasi-static model of reaming,
ASME J. Man. Sci. \& Eng., {\bf 123} (3), 387-396  (2001).

\bibitem{BAYLY3}  Bayly, P.V., M.T. Lamar, and S.G. Calvert, Low frequency
regenerative vibration and the formation of lobed holes in drilling,
ASME J. Man. Sci. \& Eng., {\bf 124} (2), 275-285  (2001).

\bibitem{Stepan} St{\'e}p{\'a}n, G.,  Delay differential equation models 
for machine tool chatter, Dynamics and Chaos in Manufacturing
Processes, F. Moon, ed., Wiley and Sons, 165-192 (1998).

\bibitem{ERNST} Ernst, H. and M.E. Merchant, Chip formation, friction and high
quality machined surfaces, Trans. Am. Soc. Met., {\bf 29}, 299-378 (1941).

\bibitem{Wellbourne} Wellbourne, D.B, and Smith, J.D., Machine Tool Dynamics, an
Introduction, Cambridge University Press, Cambridge, U.K., p. 31 (1970).  

\bibitem{STONE} Stone, E., and A. Askari,  Nonlinear models of chatter in 
drilling processes, Dynamical Systems, {\bf 17} (1) 65-85 (2002).

\bibitem{Tobias} Tobias, S.A.,
Machine Tool Vibration, J. Wiley, New York, (1965).


\end{thebibliography}
\end{document}